\documentclass[usenatbib]{mnras}
\usepackage{graphics}
\usepackage{lipsum}
\usepackage{amsmath}
\usepackage{breqn}
\usepackage{amssymb}
\usepackage{hyperref}
\usepackage{subfigure}
\usepackage{multicol}
\usepackage{caption}
\usepackage{multirow}
\usepackage{ulem}
\usepackage{rotating}
\usepackage{pdflscape}

\usepackage[utf8x]{inputenc}
\usepackage{booktabs}

\usepackage{color}

\title{Cosmic variance of $z>7$ galaxies: Prediction from \texttt{BlueTides}}
\author[Bhowmick et al.]{
Aklant K. Bhowmick$^{1}$,
Rachel S. Somerville$^{2,3}$,
Tiziana Di Matteo$^{1}$,
Stephen Wilkins$^{4}$,
\newauthor
Yu Feng $^{5}$,
Ananth Tenneti$^{1}$
\\
$^{1}$McWilliams Center for Cosmology, Dept. of Physics, Carnegie Mellon University,
Pittsburgh PA 15213, USA\\
$^{2}$Center for Computational Astrophysics, Flatiron institute, New York, NY 10010, USA\\
$^{3}$Department of Physics and Astronomy, Rutgers University, 136\\ $^{4}$Astronomy Centre, Department of Physics and Astronomy, University of Sussex, Brighton, BN1 9QH, UK\\
$^{5}$Berkeley Center for Cosmological Physics, University of California at Berkeley, Berkeley, CA 94720, USA}

\begin{document}
\maketitle
\begin{abstract}
In the coming decade, a new generation of telescopes, including JWST and WFIRST, 
will probe the period of the formation of first galaxies and quasars, and
open up the last frontier
for structure formation. Recent simulations as well as observations have suggested that these galaxies are strongly clustered~(with large scale bias $\gtrsim6$), and therefore have significant cosmic variance. In this work, we use \texttt{BlueTides}, the largest volume cosmological simulation of galaxy formation, to directly estimate the cosmic variance for current and upcoming surveys. Given its resolution and volume, \texttt{BlueTides} can probe the bias and cosmic variance of $z>7$ galaxies between magnitude $M_{UV}\sim-16$ to $M_{UV}\sim-22$ over survey areas $\sim0.1~\mathrm{arcmin}^2$ to $\sim 10~\mathrm{deg}^2$. Within this regime, the cosmic variance decreases with survey area/ volume as a power law with exponents
between $\sim-0.25$ to $\sim-0.45$. For the planned $10~\mathrm{deg}^2$ field of WFIRST, the cosmic variance is between $3\%$ to $10\%$. Upcoming JWST medium/ deep surveys with areas up to $A\sim100~\mathrm{arcmin}^2$ will have cosmic variance ranging from $\sim 20-50\%$. Lensed surveys have the highest cosmic variance $\gtrsim 40\%$; the cosmic variance of $M_{UV}\lesssim-16$ galaxies is $\lesssim100\%$ up to $z\sim11$. At higher redshifts such as $z\sim12~(14)$, effective volumes of $\gtrsim(8~\mathrm{Mpc}/h)^3$~($\gtrsim(12~\mathrm{Mpc}/h)^3$) are required to limit the cosmic variance to within $100\%$. Finally, we find that cosmic variance is larger than Poisson variance and forms the dominant component of the overall uncertainty in all current and upcoming surveys. We present our calculations in the form of simple fitting functions and an online cosmic variance calculator (\href{https://github.com/akbhowmi/CV_AT_COSMIC_DAWN}{\texttt{CV_AT_COSMIC_DAWN}}) which we publicly release.
\end{abstract}

\section{Introduction}
The underlying non-linear structure of the universe and the physics of galaxy formation are imprinted in the abundances of observable galaxies, typically characterized by the galaxy luminosity function~(LF) or stellar mass function~(SMF). Therefore, a precise measurement of the LF and SMF, and its evolution through cosmic time, is of paramount importance. 
To this end, there has been significant progress in constraining LFs and SMFs at high redshifts ~\citep{Duncan2014,2015ApJ...803...34B,Song2016,Bouwen2017,Livermore2017} using galaxies within the legacy and frontier fields of the Hubble Space Telescope as well as data from Subaru Hyper Suprime Cam. Different parts of the LF can potentially be used to probe different aspects of structure and galaxy formation. For instance, the faint end~($29\lesssim H\lesssim33$) measurements coming from lensed surveys can provide constraints on the nature of dark matter \citep{2016ApJ...818...90M,2017ApJ...836...61M,2019arXiv190401604N}. The faint end is also sensitive to modeling of stellar winds~\citep{2019MNRAS.483.2983Y}. On the other hand, the bright end is sensitive to the modeling of AGN feedback as well as dust extinction~\citep{2008MNRAS.391..481S,2015ARA&A..53...51S}.     

The next generation of infrared surveys such as JWST~\citep{2006SSRv..123..485G} and WFIRST~\citep{2015arXiv150303757S} will reach unprecedented depths, vastly increasing the sizes of high-redshift~($z>7$) galaxy samples. A major impediment in constraining the LF and SMF comes from the fact that galaxies are not uniformly distributed in space~(referred to as galaxy clustering), and therefore the number density estimates obtained from these deep~(limited in volume) surveys are susceptible to significant field-to-field variance, which cosmologists refer to as \textit{cosmic variance}\footnote{Some use the term ``cosmic variance'' to refer to the uncertainty due to our being able to probe only a limited fraction of the Universe within our cosmic horizon. Here we use the term to mean ``field-to-field'' variance}.

Recent observational measurements~\citep{2014ApJ...793...17B,2016ApJ...821..123H} have suggested that $z>7$ galaxies exhibit exceptionally strong clustering properties~(large scale galaxy bias~$>6$). This has also been predicted by recent hydrodynamic simulations~\citep{2018MNRAS.474.5393B} and semi-analytic modeling \citep{2017MNRAS.472.1995P}. Therefore, cosmic variance is expected to be a significant, potentially dominant component of the uncertainty for these high-z galaxies~(the other component being the Poisson variance arising from finite number counts).


In order to estimate the cosmic variance of a given galaxy population, the clustering strength must be known. For populations for which the clustering is well known, the cosmic variance is straightforward to compute~\citep{2004ApJ...600L.171S}. However, for the majority of galaxy populations, the clustering and galaxy bias are difficult to measure and are not well known. In such a case, several theoretical approaches may be adopted to predict the galaxy clustering. This includes clustering predictions using halo occupation models \citep{2010ApJ...710..903M,2012ApJ...752...41Y,2018MNRAS.477..359C}, semi-analytic models~\citep{2006MNRAS.369.1009B,2017MNRAS.472.1995P} and hydrodynamic simulations~\citep{2015MNRAS.450.1349K,2017MNRAS.470.1771A}.

In the recent past, clustering predictions from Halo Occupation modeling \citep{2008ApJ...676..767T,2011ApJ...731..113M} and Semi-Analytic modeling~\citep{2019ApJ...872..180C} have been used to predict the cosmic variance, each focusing on a variety of redshift regimes. \cite{2008ApJ...676..767T} in particular, analyzes the effect of cosmic variance on the shapes of luminosity functions at high redshifts~(up to $z\sim15$) by assuming an empirical one-to-one relation between halo mass and galaxy luminosity. The recent \cite{2020arXiv200411096U} uses semi-analytical modeling on dark matter only simulations and estimates the impact of reionization feedback models on the cosmic variance at $z\gtrsim6$. With \texttt{BlueTides}~\citep{2016MNRAS.455.2778F}, which is a recent cosmological hydrodynamic simulation for the high redshift universe, we now have access to the full galaxy population at $z\gtrsim7$, and are able to make ``ab initio" predictions of the galaxy clustering~\citep{2016MNRAS.463.3520W,2018MNRAS.474.5393B} and the galaxy-halo connection~\citep{2018MNRAS.480.3177B}. 
Importantly, these "ab initio" simulations naturally include scatter in the halo mass vs. galaxy luminosity relationship, based on the physical processes that shape galaxy formation in each halo, as well as the second order correlations such as assembly bias.
In this work, we use standard methodology for describing cosmic variance from the literature \citep[e.g.][]{2004ApJ...600L.171S,2008ApJ...676..767T} combined with clustering predictions from \texttt{BlueTides} to make cosmic variance estimates for the number counts and the luminosity functions for fields targeting very high redshift~($z\sim7-14$) galaxies. Section \ref{methods_sec} describes the basic methodology. Section \ref{cosmic_variance_sec} investigates the dependence of the cosmic variance on the various survey parameters, and also summarizes the cosmic variance estimates for the planned deep fields of JWST and WFIRST. We provide our main conclusions in Section \ref{summary_sec}.

\section{Methods}
\label{methods_sec}
\subsection{\texttt{BlueTides} Simulation}

\texttt{BlueTides} is a high resolution
cosmological hydrodynamic simulation run until $z\sim 7.5$ using the cosmological code \texttt{MP-GAGDET}. With a
simulation box size of $(400~\mathrm{Mpc}/h)^{3}$ and $2\times
7048^{3}$ particles, \texttt{BlueTides} has a resolution comparable to
\texttt{Illustris}~\citep{Nelson201512},
\texttt{Eagle}~\citep{2015MNRAS.446..521S},
\texttt{MassiveBlackII}~\citep{2015MNRAS.450.1349K} but is $\sim64$
times the volume. The cosmological parameters are derived from the nine-year Wilkinson Microwave Anisotropy Probe (WMAP)
\citep{2013ApJS..208...19H} ($\Omega_0=0.2814$,
$\Omega_\lambda=0.7186$, $\Omega_b=0.0464$ $\sigma_8=0.82$, $h=0.697$,
$n_s=0.971$). The dark matter and gas particles
  have masses $1.2\times10^{7}~M_{\odot}/h$,
  $2.36\times10^{6}~M_{\odot}/h$ respectively.  
We identify haloes using an FOF Group finder
\citep{1985ApJ...292..371D}, and the halo substructure
using \texttt{ROCKSTAR-GALAXIES} \citep{2013ApJ...762..109B}.  For
more details on \texttt{BlueTides}, interested readers should refer to
\cite{2016MNRAS.455.2778F}.

The various sub-grid physics models that have been employed in \texttt{BlueTides} include a multiphase model for star formation  \citep{2003MNRAS.339..289S,2013MNRAS.436.3031V},
Molecular hydrogen formation \citep{2011ApJ...729...36K},
gas and metal cooling \citep{1996ApJS..105...19K,2014MNRAS.444.1518V}, SNII feedback \citep{Nelson201512}, Black hole growth and AGN feedback \citep{2005MNRAS.361..776S,2005Natur.433..604D}, and a model for ``Patchy" reionization \citep{2013ApJ...776...81B}.

\texttt{BlueTides} was targeted towards the high redshift~($z>7$) Universe, with its large volume that captures the statistics of the brightest~(rarest) galaxies and quasars. 
The UV luminosity functions \citep{2015ApJ...808L..17F, 2016MNRAS.455.2778F,
2016MNRAS.463.3520W} are consistent with existing observational constraints \citep{2015ApJ...803...34B}. In addition, the predictions are broadly consistent across different hydrodynamic simulations and semi-analytic models~\citep{2019MNRAS.483.2983Y,2019arXiv190105964Y}. Clustering properties are also consistent with currently available observations \citep{2018MNRAS.474.5393B}. \texttt{BlueTides} has also enabled us to build Halo Occupation Distribitions ~(HOD) models for clustering of galaxies in the of $z>7.5$ regime~\citep{2018MNRAS.480.3177B}. Photometric properties of high redshift galaxies and the effect of stellar population synthesis modeling as well as dust modeling have been extensively studied in \cite{2016MNRAS.458L...6W, 2016MNRAS.460.3170W, 2018MNRAS.473.5363W}. \texttt{BlueTides} has allowed the study of the rare earliest supermassive black holes/first quasars and the role of tidal field in the black hole growth in the early universe \citep{2017MNRAS.467.4243D}. Dark matter only realizations have been used to trace their descendents to the present day \citep{2017arXiv170803373T}. We have also been able to make predictions from \texttt{BlueTides} \citep{2019MNRAS.483.1388T, 2018MNRAS.481.4877N} for the recently discovered highest redshift quasar \citep{2018Natur.553..473B}.


The galaxy spectral energy distributions (SEDs) were calculated using
the \texttt{PEGASE-v2} \citep{1997A&A...326..950F} stellar population synthesis (SPS) models with the 
stellar initial mass function of \cite{2003PASP..115..763C}. The cumulative SED for each galaxy is the sum of the SEDs for each star particle (as a function of stellar age and metallicity). For a complete discussion of the photometric properties of \texttt{BlueTides} galaxies we urge the readers to refer to \cite{2016MNRAS.460.3170W}. 

We shall be considering galaxy samples limited by a $M_{UV}$ band absolute magnitude~(denoted by $M_{UV}(<)$), which corresponds to $1600~\mathrm{A}$ in the rest frame SED of the galaxies. Given its high resolution as well as large volume, \texttt{BlueTides} is able to probe the clustering, and therefore the cosmic variance of galaxies with magnitudes ranging from $M_{UV}\sim-16$ to $M_{UV}\sim-22$. Hereafter, we shall discuss the cosmic variance of galaxies within this magnitude range unless stated otherwise. Note that we do not include a dust correction in the calculation of the magnitudes since its effect is significant only at the very bright end~($M_{UV}\sim-22$ to $\sim-25$)~\citep[Figure. 10]{2016MNRAS.455.2778F}.    
\subsection{Determining cosmic variance}
\label{cosmic_variance_method}
The number of objects $N$ within a field of view with volume $V$ can be described by a probability distribution $P(N|V)$. The cosmic variance~($\sigma_g$) can then be defined as 
\begin{equation}
\sigma_g^2=\frac{\left<N^2\right>-\left<N\right>^2-\left<N\right>}{\left<N\right>^2}
\label{eqn_definition}
\end{equation}
where the $p^{\mathrm{th}}$ moment of $P(N|V)$ is given by $\left<N^p\right>=\sum_N N^p P(N|V)$. The first two terms in Eq.~(\ref{eqn_definition}) represent the total variance in $N$ which includes the contribution from cosmic variance and Poisson variance. The third term represents the Poisson variance which is subtracted to obtain $\sigma_g^2$.

We use the \texttt{BlueTides} simulation to determine $\sigma_g^2$ by computing the two-point galaxy correlation function~$\xi_{gg}$ of \texttt{BlueTides} galaxies and integrating it over the relevant volume, as in \citet[page 234]{1980lssu.book.....P}. $\sigma_g^2$ can calculated using
\begin{equation}
\sigma_g^2=\frac{1}{V^2}\int_{V} \xi_{gg}(\mathbf{r_1},\mathbf{r_2}) d^3\mathbf{r_1} d^3\mathbf{r_2}
\label{xi_to_sigma}
\end{equation} 
where $\mathbf{r_1}$ and $\mathbf{r_2}$ are position vectors of galaxies integrated over the survey volume. With this approach, we can determine the cosmic variance for survey volumes as large as the \texttt{BlueTides} volume. In addition, for survey volumes~(e.g. JWST medium / deep surveys and lensed surveys) that are small enough such that a sufficiently large number of them can be extracted from the simulation box, we also determine the full distribution of number counts and analyse the cosmic variance.   


We extract a mock survey volume~(corresponding to survey Area $A$ and redshift width $\Delta z$) from a single snapshot of \texttt{BlueTides}, with median redshift $z_{\rm med}$.  The survey volume $V$ is modeled as a cuboidal box with line-of-sight length determined by the comoving distance between $z \pm \Delta z/2$, and transverse dimensions given by the comoving length subtended by the survey angular size $\sqrt(A)$ at the median redshift.

\section{Cosmic variance of \texttt{BlueTides} galaxies}
\label{cosmic_variance_sec}
\subsection{Clustering of \texttt{BlueTides} galaxies}
\begin{figure*}
\includegraphics[width=\textwidth]{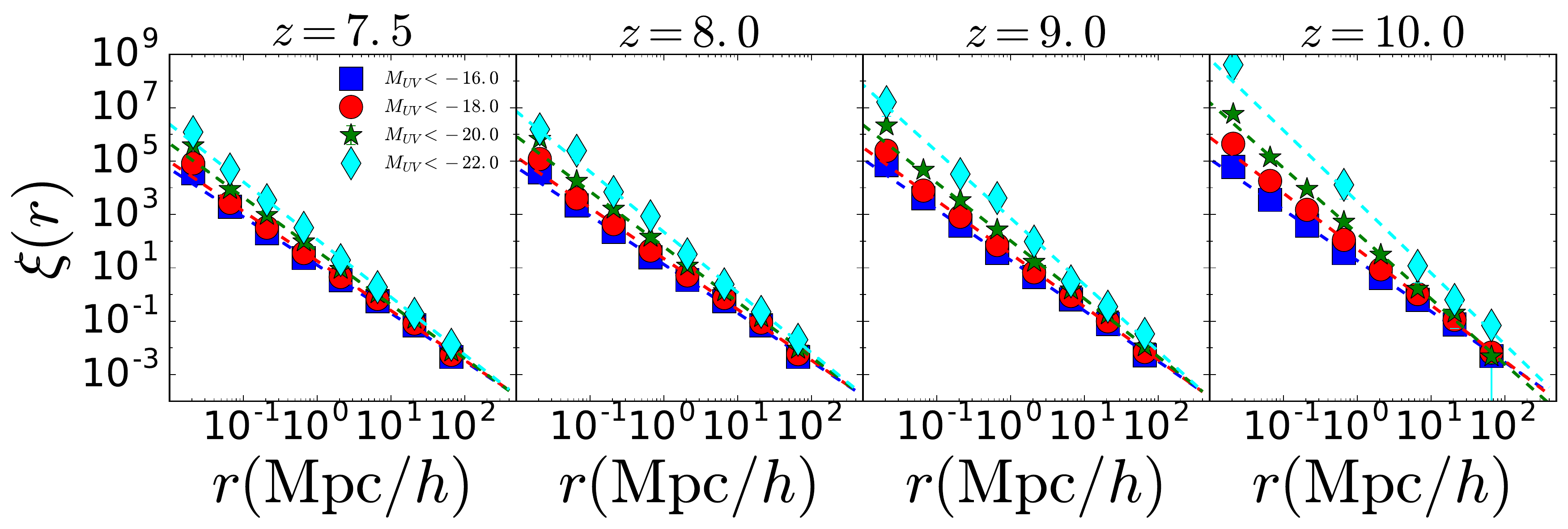}\\
\includegraphics[width=\textwidth]{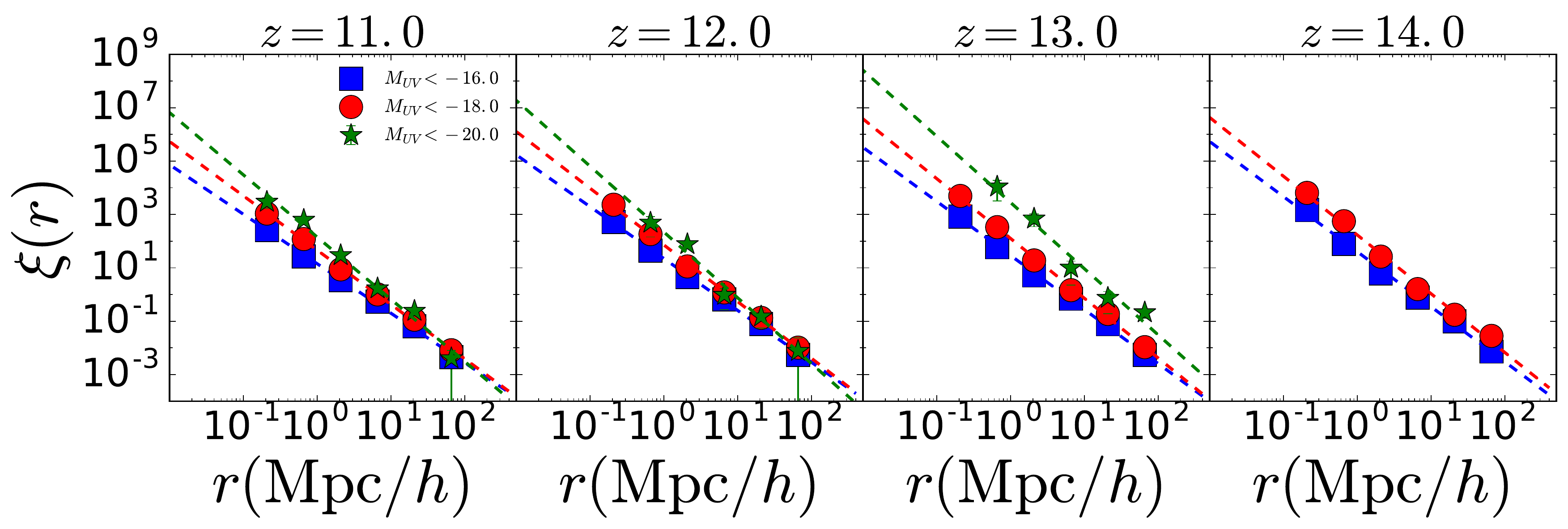}
\caption{Two-point correlation functions (circles) and their power law fits (lines) for \texttt{BlueTides} galaxies as a function of pairwise comoving distance $r$. The lines with different colors represent different $M_{UV}$ thresholds.}
\label{correlation}
\end{figure*}
Cosmic variance depends sensitively on how strongly clustered the galaxy population under consideration is; we therefore begin by presenting the clustering power of \texttt{BlueTides} galaxies. Figure \ref{correlation} shows the two-point correlation functions $\xi(r)$ of galaxies from $r\sim0.01~\mathrm{Mpc}/h$ to $r\sim400~\mathrm{Mpc}/h$. $\xi(r)$ increases with 1) decreasing $M_{UV}$ thresholds~(increasing luminosity) at fixed redshift, and 2) increasing redshift at fixed $M_{UV}$ threshold. We note that $\xi(r)$ can be well described by a power-law profile described as
\begin{equation}
\xi(r)=(r/r_0)^{\gamma}
\end{equation}
where $r_0$ is the correlation length and $\gamma$ is a power law exponent. The dashed lines in Figure \ref{correlation} show the power law fits and the corresponding best fit parameters are listed in Table \ref{power_law_fits_table}. We shall hereafter use these power-law fits to compute the cosmic variance using Eq.~(\ref{xi_to_sigma}).
\subsection{Dependence of cosmic variance on survey geometry}
Here, we compute the cosmic variance $\sigma_g$ and study its dependence on the various parameters of the survey.
\subsubsection{Survey Area}
\label{redshift_area_sec}
\begin{figure*}
\includegraphics[width=\textwidth]{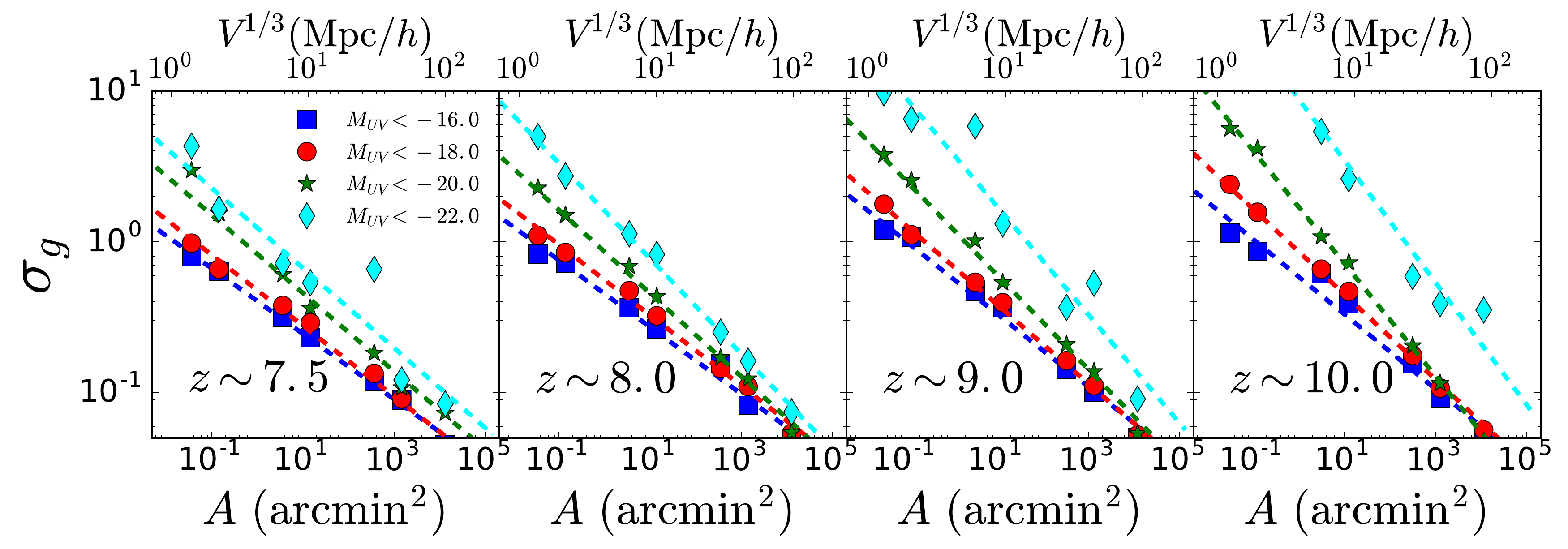}\\
\includegraphics[width=\textwidth]{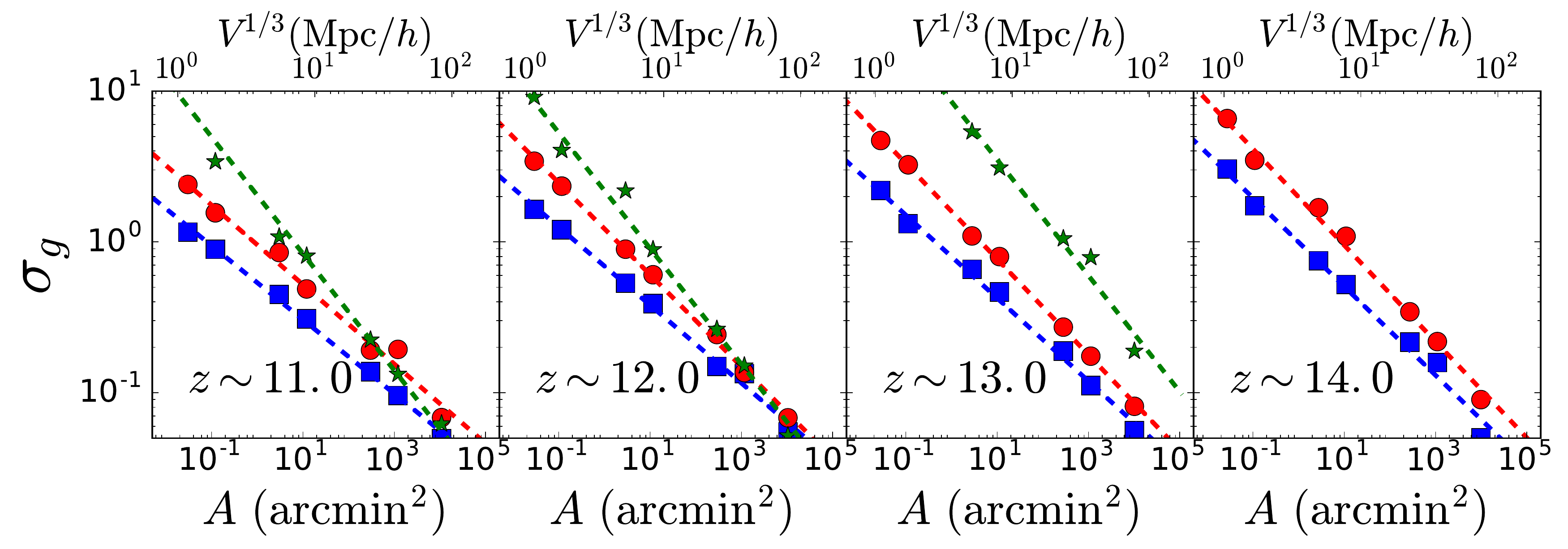}
\caption{The filled circles show the cosmic variance as a function of survey area $A$ and a redshift width of $\Delta z=1$ for various $M_{UV}$ threshold samples. Dashed lines of corresponding color show power law fits.}
\label{survey_area_fig}
\end{figure*}
Figure \ref{survey_area_fig} shows the cosmic variance as a function of survey area. Over areas ranging from $\sim1~\mathrm{arcmin}^2$ to $\sim1~\mathrm{deg}^2$, the cosmic variance can range from $\sim1-2\%$ up to $\gtrsim100\%$ depending on the magnitudes and redshifts of the galaxies. In the next section, we shall discuss in more detail the expected cosmic variance of upcoming surveys. 

The dependence of cosmic variance on survey area can be described as a power-law,
\begin{equation}
    \sigma_g=\Sigma A^\beta
\end{equation}
where $\alpha$ is the power-law exponent and $\Sigma$ is the pre-factor. This is not surprising as the clustering profile of these galaxies could also be described by a power-law. The best fit values of $\Sigma$ and $\beta$ obtained from our results are summarized in Table \ref{power_law_fits_table}.

We also investigate the dependence on the survey aspect ratio. We report no significant variation of the cosmic variance over aspect ratios ranging from $0.2$ to $1$ for fixed survey area. However, \cite{2011ApJ...731..113M} showed that for very elongated geometries~(survey aspect ratio $<0.1$), the cosmic variance can be reduced by factors $\sim5$. This is due to a larger mean distance between two galaxies detected in such a survey. For a detailed discussion we refer readers to \cite{2011ApJ...731..113M}.
\begin{table}
\begin{tabular}{|c|c|c|c|c|c|}
$M_{UV}(<)$ &$z$ & $\gamma$ & $r_0 (\mathrm{Mpc}/h)$ & $\alpha$ & $\Sigma$\\
\hline
-16 & 7.5 & -1.80 & 4.07  & -0.43 & 0.61 \\
-18 & 7.5 & -1.86 & 4.74 & -0.46 & 0.75 \\
-20 & 7.5 & -2.01 & 6.44 & -0.51 & 1.37 \\
-22 & 7.5 & -2.15 & 9.00 & -0.53 & 2.06 \\
\hline
-16 & 8.0 & -1.82 & 4.15 & -0.44 & 0.70 \\
-18 & 8.0 & -1.90 & 5.08 & -0.46 & 0.88 \\
-20 & 8.0 & -2.07 & 7.31  & -0.55 & 1.48 \\
-22 & 8.0 & -2.26 & 10.83 & -0.63 & 2.99 \\
\hline
-16 & 9.0 & -1.89 & 4.76 & -0.48 & 0.94 \\
-18 & 9.0 & -2.98 & 5.86 & -0.52 & 1.19 \\
-20 & 9.0 & -2.17 & 8.45  & -0.62 & 2.33 \\
-22 & 9.0 & -2.49 & 14.06 & -0.71 & 8.14 \\
\hline
-16 & 10.0 & -1.90 & 4.64 & -0.49 & 0.98 \\
-18 & 10.0 & -2.09 & 6.47 & -0.56 & 1.55 \\
-20 & 10.0 & -2.45 & 8.58  & -0.74 & 3.70 \\
-22 & 10.0 & -2.67 & 19.98 & -0.81 & 20.59 \\
\hline
-16 & 11.0 & -1.85 & 4.18 & -0.49 & 0.90 \\
-18 & 11.0 & -2.03 & 6.54 & -0.52 & 1.66 \\
-20 & 11.0 & -2.33 & 8.36  & -0.77 & 4.61 \\
\hline
-16 & 12.0 & -1.94 & 4.93 & -0.51 & 1.20 \\
-18 & 12.0 & -2.12 & 7.47 & -0.60 & 2.35 \\
-20 & 12.0 & -2.33 & 8.89  & -0.77 & 5.00 \\
\hline
-16 & 13.0 & -2.02 & 5.26 & -0.55 & 1.45 \\
-18 & 13.0 & -2.24 & 8.64 & -0.63 & 3.18 \\
-20 & 13.0 & -2.48 & 24.5  & -0.77 & 19.72 \\
\hline
-16 & 14.0 & -2.06& 5.98 & -0.58 & 1.90 \\
-18 & 14.0 & -2.20& 10.2 & -0.64 & 4.01 \\
\hline
\end{tabular}

\caption{Best fit values of the power law fit parameters for $\xi$ and $\sigma_g$ for galaxy samples with various $M_{UV}$ thresholds and redshifts.}
\label{power_law_fits_table}
\end{table}

\subsubsection{Redshift bin width}
\label{redshift_width_sec}

\begin{figure}
\includegraphics[width=8cm]{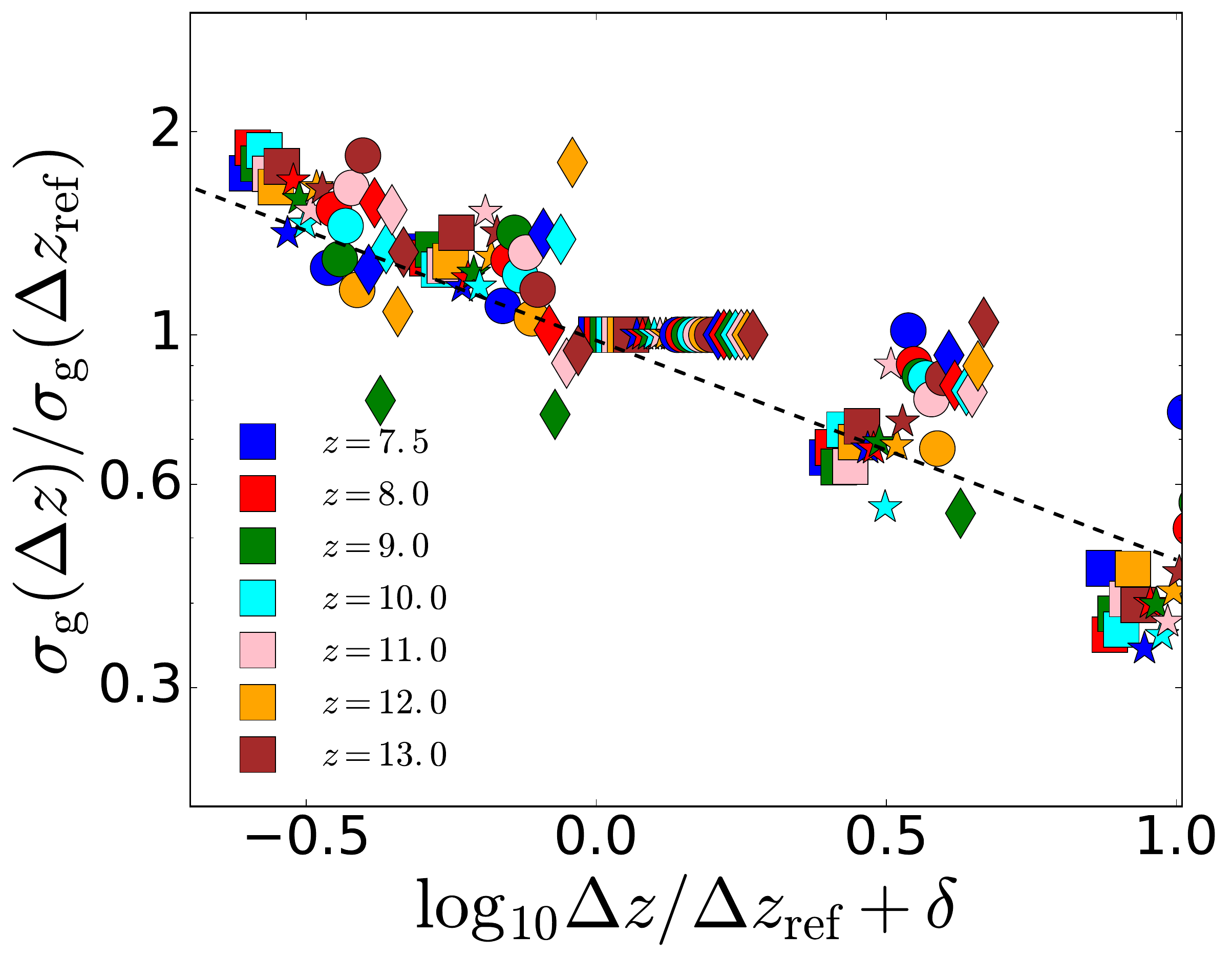}
\caption{Cosmic variance as a function of redshift bin width $\Delta z$ normalized with respect to a reference redshift width $\Delta z_{\mathrm{ref}}=1$. We show this for galaxies with $M_{UV}<-16$. $\delta$ is a small~($<0.1$) horizontal offset added to the $x$ axis to avoid overlap between the data points. The black dashed line corresponds to the best fit power-law. Circles and stars correspond to survey areas of $10~\mathrm{deg}^2$ and $1~\mathrm{deg}^2$ respectively. Squares correspond to survey area of $10~\mathrm{arcsec}^2$.}
\label{redshift_width}
\end{figure}
Figure \ref{redshift_width} shows the dependence of $\sigma_g$ on redshift bin width for $M_{UV}<-16$ galaxies. As expected, $\sigma_g$ decreases as $\Delta z$ increases due to the increase in the comoving volume of the survey. Furthermore, the ratio $\sigma_g(\Delta z)/\sigma_g(\Delta z_{\mathrm{ref}})$~(where the reference redshift width $\Delta z_{\mathrm{ref}}$ is chosen to be 1 in Figure \ref{redshift_width}) has a somewhat universal power-law dependence on $\Delta z$, independent of magnitude, redshift and survey type. This behavior is also reported for $z<3$ galaxies~\citep{2011ApJ...731..113M}. We determine the best fit power-law~(shown as the black dashed line) to be
\begin{equation}
\sigma_g(\Delta z)/\sigma_g(\Delta z_{\mathrm{ref}})=(\Delta z/\Delta z_{\mathrm{ref}})^{-0.32}.
\label{redshift_width_variation_eqn}
\end{equation}  

\subsection{Dependence of cosmic variance on galaxy UV magnitude}
\begin{table}
\begin{tabular}{|c|c|c|c|}
Survey & $z$ & $M_{UV}(<)$ & $\left< N \right> \pm \delta N_{\mathrm{cosmic}} \pm \delta N_{\mathrm{Poisson}}$ \\
\hline
\hline
$10~\mathrm{arcmin}^2$ & $7.5$ & $-16$ & $210.9\pm63.6\pm14.5$ \\
$10~\mathrm{arcmin}^2$ & $7.5$ & $-18$ & $45.3\pm17.2\pm6.7$ \\
$10~\mathrm{arcmin}^2$ & $7.5$ & $-20$ & $7.7\pm3.9\pm2.8$ \\
\vspace{0.3cm}
$10~\mathrm{arcmin}^2$ & $7.5$ & $-22$ & $0.9\pm0.6\pm0.9$ \\

$10~\mathrm{arcmin}^2$ & $11.0$ & $-16$ & $14.3\pm7.0\pm3.8$ \\

$10~\mathrm{arcmin}^2$ & $11.0$ & $-18$ & $1.0\pm0.8\pm1.0$ \\
\vspace{0.3cm}
$10~\mathrm{arcmin}^2$ & $11.0$ & $-20$ & $0.0\pm0.1\pm0.2$ \\

$10~\mathrm{arcmin}^2$ & $14.0$ & $-16$ & $0.8\pm0.6\pm0.9$ \\
$10~\mathrm{arcmin}^2$ & $14.0$ & $-18$ & $0.0\pm0.0\pm0.2$ \\
\hline
\hline
$100~\mathrm{arcmin}^2$ & $7.5$ & $-16$ & $2182.2\pm451.0\pm46.7$ \\
$100~\mathrm{arcmin}^2$ & $7.5$ & $-18$ & $469.7\pm118.8\pm21.7$ \\
$100~\mathrm{arcmin}^2$ & $7.5$ & $-20$ & $79.8\pm25.7\pm8.9$ \\
\vspace{0.3cm}
$100~\mathrm{arcmin}^2$ & $7.5$ & $-22$ & $9.4\pm4.2\pm3.1$ \\

$100~\mathrm{arcmin}^2$ & $11.0$ & $-16$ & $148.9\pm45.4\pm12.2$ \\
$100~\mathrm{arcmin}^2$ & $11.0$ & $-18$ & $10.9\pm4.5\pm3.3$ \\
\vspace{0.3cm}
$100~\mathrm{arcmin}^2$ & $11.0$ & $-20$ & $0.5\pm0.3\pm0.7$ \\

$100~\mathrm{arcmin}^2$ & $14.0$ & $-16$ & $8.5\pm3.7\pm2.9$ \\
$100~\mathrm{arcmin}^2$ & $14.0$ & $-18$ & $0.3\pm0.2\pm0.5$ \\
\hline
\hline
$1~\mathrm{deg}^2$ & $7.5$ & $-16$ & $79210.7\pm4585.6\pm281.4$ \\
$1~\mathrm{deg}^2$ & $7.5$ & $-18$ & $17058.8\pm1220.6\pm130.6$ \\
$1~\mathrm{deg}^2$ & $7.5$ & $-20$ & $2912.1\pm241.3\pm54.0$ \\
\vspace{0.3cm}
$1~\mathrm{deg}^2$ & $7.5$ & $-22$ & $340.2\pm37.1\pm18.4$ \\

$1~\mathrm{deg}^2$ & $11.0$ & $-16$ & $5386.6\pm531.6\pm73.4$ \\
$1~\mathrm{deg}^2$ & $11.0$ & $-18$ & $391.8\pm57.1\pm19.8$ \\
\vspace{0.3cm}
$1~\mathrm{deg}^2$ & $11.0$ & $-20$ & $16.6\pm4.1\pm4.1$ \\

$1~\mathrm{deg}^2$ & $14.0$ & $-16$ & $305.5\pm41.0\pm17.5$ \\
$1~\mathrm{deg}^2$ & $14.0$ & $-18$ & $10.6\pm2.1\pm3.3$ \\
\hline
\hline
$10~\mathrm{deg}^2$ & $7.5$ & $-16$ & $791981\pm23759\pm889$ \\
$10~\mathrm{deg}^2$ & $7.5$ & $-18$ & $170563\pm6822\pm413$ \\
$10~\mathrm{deg}^2$ & $7.5$ & $-20$ & $29079\pm1454\pm171$ \\
\vspace{0.3cm}
$10~\mathrm{deg}^2$ & $7.5$ & $-22$ & $3368\pm168\pm58$ \\

$10~\mathrm{deg}^2$ & $11.0$ & $-16$ & $53595\pm1608\pm231$ \\
$10~\mathrm{deg}^2$ & $11.0$ & $-18$ & $3910\pm117\pm62$ \\
\vspace{0.3cm}
$10~\mathrm{deg}^2$ & $11.0$ & $-20$ & $168\pm7\pm13$ \\

$10~\mathrm{deg}^2$ & $14.0$ & $-16$ & $3019\pm121\pm54$ \\
$10~\mathrm{deg}^2$ & $14.0$ & $-18$ & $102\pm4\pm10$ \\
\hline
\hline
\end{tabular}

\caption{Average number of galaxies $\left<N\right>$ at various threshold UV magnitudes~$M_{UV}~(<)$ and redshifts along with their uncertainties due to cosmic variance~($\delta N_{\mathrm{cosmic}}$) and Poisson variance~($\delta N_{\mathrm{poisson}}$). The survey areas correspond to those presented in Figure \ref{stellar_mass_threshold}.}
\label{number_counts_table}
\end{table}

\begin{figure*}
\includegraphics[width=\textwidth]{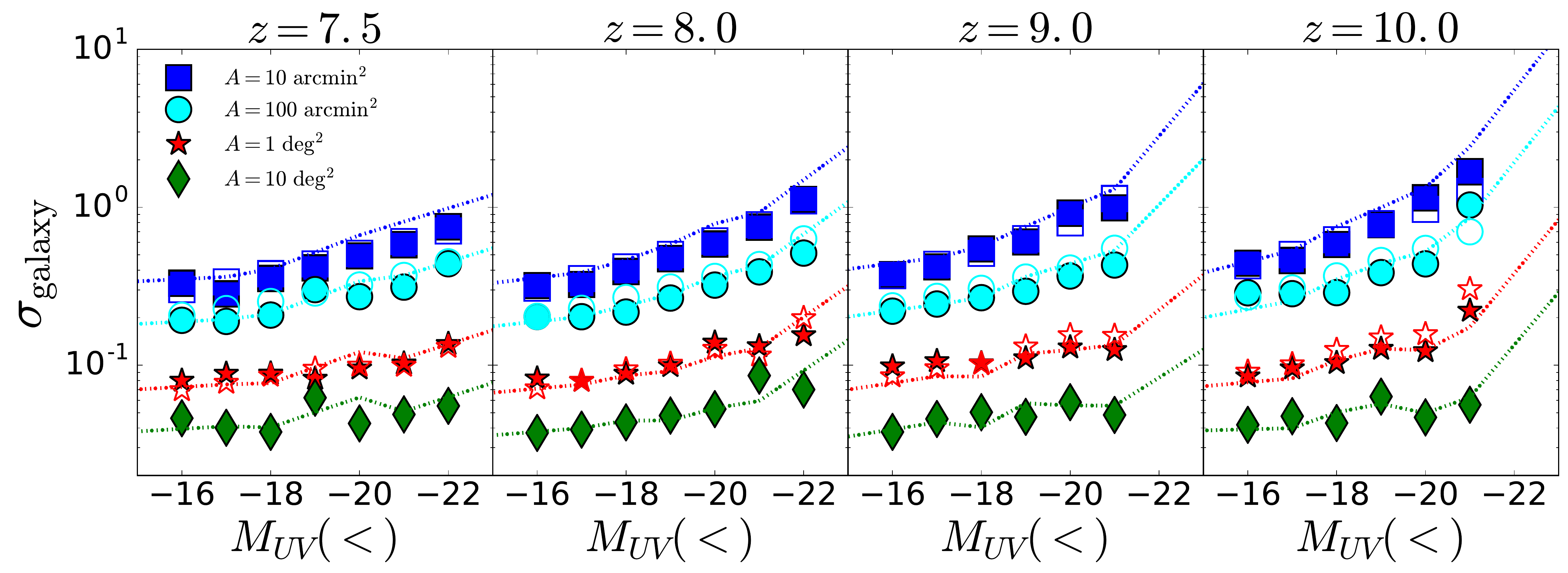}\\

\includegraphics[width=\textwidth]{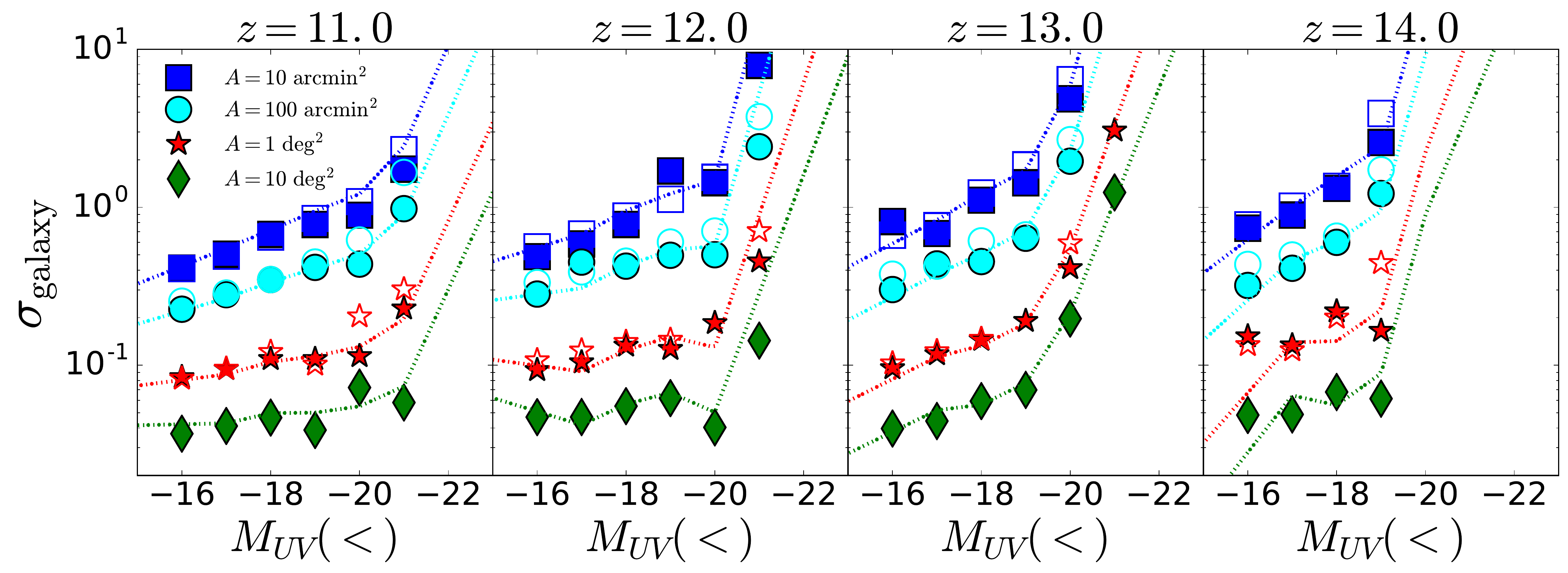}\\
\caption{Filled circles show the cosmic variance as a function of $M_{UV}$ threshold for various survey areas~(A) with $\Delta z=1$. The filled data points are computed by integrating the correlation function. The open data points for $A=1~\mathrm{arcmin}^2,10~\mathrm{arcmin}^2$,$1~\mathrm{\mathrm{deg}}^2$ are computed from the full distribution of number counts of galaxies for an ensemble of simulation sub-volumes~(representing survey volumes). The dotted lines are estimates provided by \href{https://github.com/akbhowmi/CV_AT_COSMIC_DAWN}{\texttt{CV_AT_COSMIC_DAWN}}}.
\label{stellar_mass_threshold}
\end{figure*}

\begin{figure}
\includegraphics[width=8.5 cm]{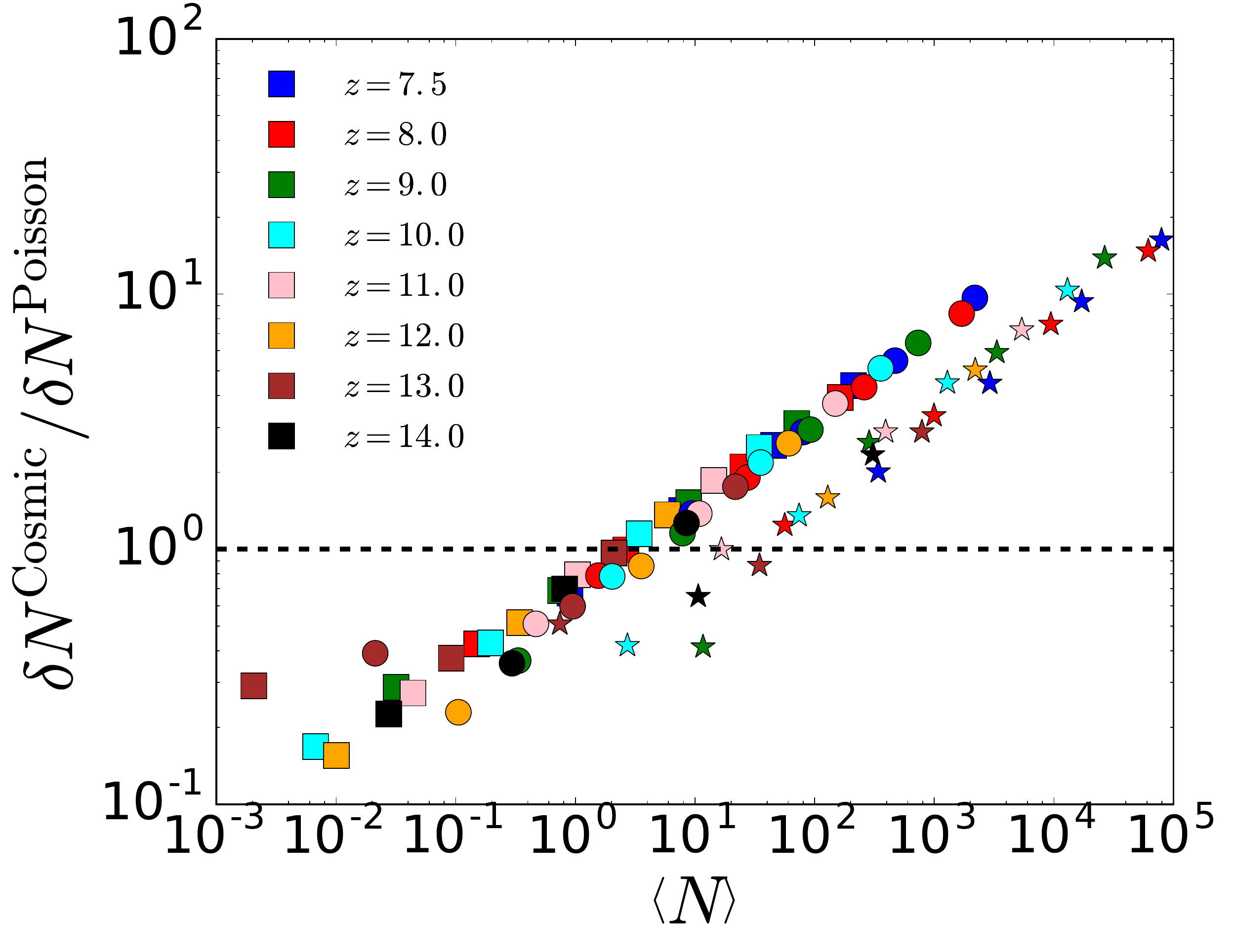}\\
\caption{$\left<N\right>$ is the mean value of the number count of galaxies in a  survey. $\left<\delta N^{\mathrm{Cosmic}}\right>/\left<\delta N^{\mathrm{Poisson}}\right>$ is the ratio between the uncertainties in the number counts contributed by cosmic variance vs. Poisson variance. The squares, circles and star shaped markers correspond to survey areas of $10~\mathrm{arcmin}^2$, $100~\mathrm{arcmin}^2$ and $1~\mathrm{deg}^2$ respectively. The redshift width is assumed to be 1. These numbers are computed from the full distribution of number counts for an ensemble of simulation sub-volumes.}
\label{cosmic_vs_poisson}
\end{figure}

We now investigate the dependence of cosmic variance on absolute UV magnitude. We shall present results for survey geometries most relevant to upcoming deep surveys within JWST and WFIRST. They also cover a wide range of existing surveys which are listed in Table \ref{list_of_surveys_table}  

\subsubsection{JWST and WFIRST-like volumes}
Figure \ref{stellar_mass_threshold} shows the cosmic variance $\sigma_g$ as a function of $M_{UV}$ threshold at the redshift snapshots $7.5-14$. The areas are representative of planned WFIRST~($1-10~\mathrm{deg}^2$) deep surveys as well as JWST~($10-100~\mathrm{arcmin}^2$) medium/ deep surveys such as JADES and CEERS survey. We show redshift widths $\Delta z\sim1$ as the photometric redshift uncertainties are expected to be significant. For the $1,10~\mathrm{arcmin}^2$ and $1~\mathrm{deg}^2$ survey, we also present the estimates from the full distribution of number counts for an ensemble of simulation sub-volumes; we find that these estimates are in reasonable agreement with those computed by integrating the correlation functions.   

We see that the cosmic variance increases with decreasing $M_{UV}$  at fixed redshift, which is expected since brighter galaxies are more strongly clustered~\citep{2017MNRAS.472.1995P, 2018MNRAS.474.5393B,2018MNRAS.480.3177B}. The scaling of the cosmic variance with respect to $M_{UV}$ can range from $\sim |M_{UV}|^{2}$ to $\sim |M_{UV}|^{4}$ for $M_{UV}$ between $-16$ to $-20$. For the more luminous galaxies with $M_{UV}$ between $-20$ to $-22$, the scaling is steeper, particularly at $z\sim10-14$. The redshift dependence~(at fixed UV magnitude) of the cosmic variance is driven by the evolution of the galaxy clustering~(increases with redshift) as well as comoving survey volume~(increases with redshift for fixed survey geometry), wherein the former tends to increase and the latter tends to decrease the cosmic variance; for $M_{UV}$ between $-16$ to $-20$ the two effects roughly cancel each other, leading to very marginal redshift dependence. For $M_{UV}<-20$, the cosmic variance increases with redshift since the clustering evolution becomes more pronounced, and becomes more important than the redshift dependence of the comoving survey volume. 

We now broadly summarize the cosmic variance predictions for the various survey areas. For the  $10~\mathrm{deg}^2$ and $1~\mathrm{deg}^2$ fields spanning the areas for the planned WFIRST deep surveys, the cosmic variance is $\sim3-10\%$ for the entire range of $M_{UV}$ at $z=7.5-9$; at higher redshifts~($z\sim9-14$), the cosmic variance is $\sim3-10\%$ for $M_{UV}$ between $-16$ to $-20$, but can exceed $\sim10\%$ for galaxies with $M_{UV}<-20$. The cosmic variance is significantly higher for $100~\mathrm{arcmin}^2$ and $10~\mathrm{arcmin}^2$ fields which span the areas for upcoming JWST medium/ deep surveys. For a $100~\mathrm{arcmin}^2$ field, the cosmic variance is between $20-50\%$ for UV magnitudes between $-16$ and $-20$. For a $10~\mathrm{arcmin}^2$ field, the cosmic variance ranges from $30-70\%$ for $M_{UV}$ between $-16$ and $-18$ up to $z\sim12$. For $z>12$, the cosmic variance within a $10~\mathrm{arcmin}^2$ field is $\gtrsim100\%$ for the entire range of UV magnitudes between $-16$ to $-22$. 

We now cast these results in terms of the overall uncertainties in the expected number counts, which are summarized in Table \ref{number_counts_table}. In a $10~\mathrm{deg}^2$ survey within WFIRST, we predict $\sim50,000$ galaxies at $z\sim7.5$ up to depths of $H<27.5$~($M_{UV}\lesssim-19.6$) wherein the uncertainty due to cosmic variance amounts to $\sim\pm2500$ galaxies; at $z\sim11$, we expect $\sim170\pm7$ galaxies. In a $1~\mathrm{deg}^2$ survey within WFIRST, we predict $\sim10,000\pm800$ galaxies at $z\sim7.5$ up to depths of $H<28.5$~($M_{UV}\lesssim-18.6$); at $z\sim11$, we predict $\sim85\pm10$ galaxies. For an area of $\sim100~\mathrm{arcmin}^2$ which broadly represents the JADES-deep and CEERS surveys, we expect to detect $\sim2200\pm450$ galaxies with $M_{UV}<-16$ at $z\sim7.5$; likewise, at $z\sim11$, we expect to detect $\sim150\pm45$ galaxies.

Lastly, we also look at the relative importance between the uncertainty due to cosmic variance and Poisson variance. In Figure \ref{cosmic_vs_poisson}, we show the ratio between cosmic variance and the Poisson variance, as a function of the number counts of galaxies for a range of redshifts, $M_{UV}$ thresholds and survey areas. We see that as we increase the number counts, the importance of cosmic variance~(relative to Poisson variance) increases. Additionally, at fixed number count, a $1~\mathrm{deg}^2$ survey~(star shaped points) has a lower cosmic variance~(relative to Poisson variance) compared to the $10,100~\mathrm{arcmin}^2$ surveys. Most importantly, we find that~(apart from the obvious exception of very low galaxy number counts i.e. $N\lesssim10$), the uncertainty due to cosmic variance largely dominates over the Poisson variance.


\noindent \textit{Encountering rare luminous galaxies and environments in upcoming JWST medium / deep surveys:}

\begin{figure}
\includegraphics[width=8 cm]{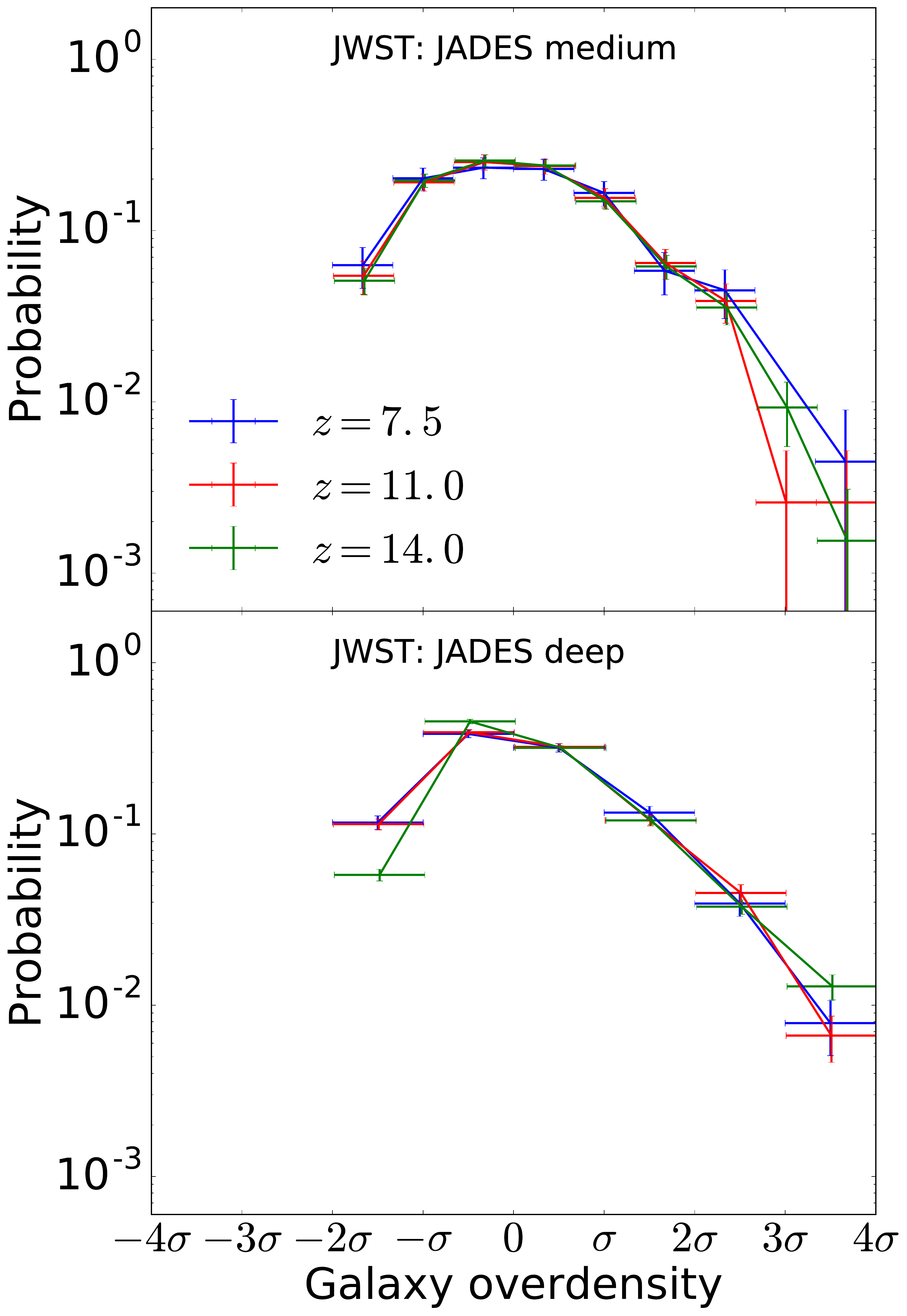}
\caption{The probability distribution of galaxy overdensities~(in the units of the standard deviation $\sigma$ of the distributions) within the ensemble of subvolumes corresponding to JWST fields~(JADES-medium and JADES-deep). The three different colors correspond to redshifts~7.5,11 and 14 spanning the entire range of interest. These are made for galaxies with $M_{UV}<-16$, but the distributions do not vary significantly for $M_{UV}$ thresholds between -16 to -19. The redshift width has been assumed to be 1.}
\label{outlier_overdensities}
\end{figure}

For the JWST medium / deep surveys~(CEERS, JADES-medium/deep), the \texttt{BlueTides} volume is large enough to produce $\gtrsim1000$ realizations. This enables us to construct the full distribution of the predicted number counts of galaxies within these surveys~(in addition to the mean and cosmic variance, which only provides the 1st and 2nd moments of the underlying distributions). We are therefore also able to probe the likelihood of encountering extreme~(several standard deviations away from the mean) overdense/ underdense regions within these surveys. Figure \ref{outlier_overdensities} shows the normalized probability distributions of overdensity of galaxies. Here, we choose to show galaxies with $M_{UV}<-16$, but the distributions~(when presented in the units of the standard deviations) do not significantly change for $M_{UV}$ between $-16$ to $-19$~(approximate range of detection limits of the JWST medium / deep surveys). Also note that there is no significant redshift evolution of these distributions~(when presented in the units of the standard deviations). The distributions of JADES deep~(lower panel) are slightly broader than that of JADES medium~(upper panel); this is expected due to lower volume of JADES deep compared to JADES medium. We find that the likelihood of these surveys to fall on underdense/ overdense regions $2\sigma$'s away from the mean, is $\sim5-10\%$. \texttt{BlueTides} contains no underdense~(void) regions with densities lower than $2\sigma$'s away from the mean. On the other hand, the most overdense regions found in \texttt{BlueTides}, correspond to $\sim3\sigma$'s away from the mean; the likelihood of these extremely overdense regions to be encountered by a JWST medium / deep survey, is about $0.1-1\%$.             

\begin{figure}
\includegraphics[width=8.5cm]{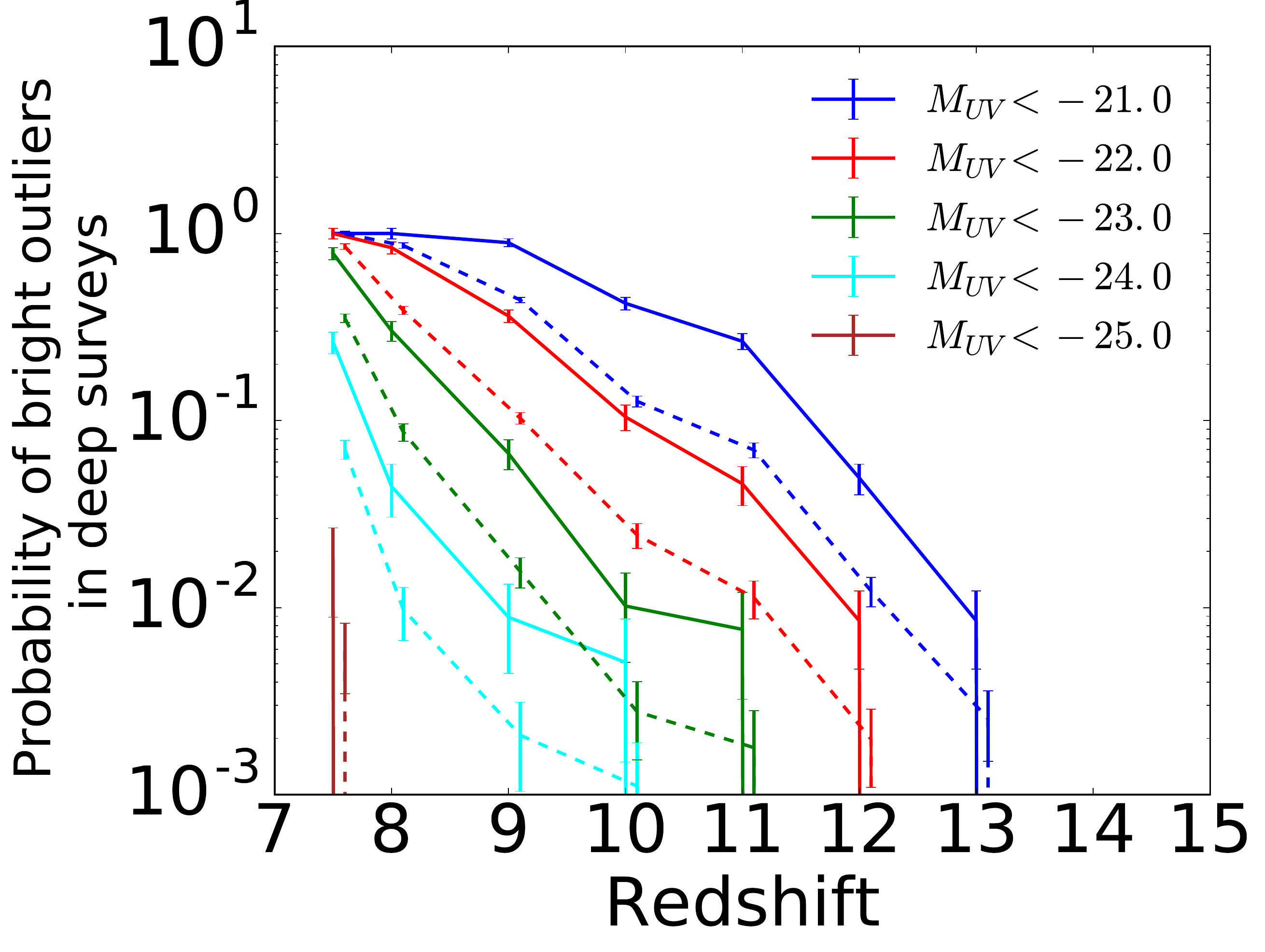}
\caption{The detection probability of bright/ luminous~($M_{UV}<-21$) galaxies in current and upcoming JWST surveys. The solid lines and dashed lines correspond to JADES-medium and the JADES-deep survey respectively. The redshift width has been assumed to be 1. $M_{UV}<-22$~(red lines) corresponds to the GNz11~\protect\citep{2016ApJ...819..129O} type galaxies. The redshift width has been assumed to be 1.}
\label{outlier_galaxies}
\end{figure}

The large volume of \texttt{BlueTides} also allows us to probe the likelihood of a~(chance) detection of rare luminous~($M_{UV}<-22$) galaxies within the JWST medium / deep surveys. We had so far not discussed these objects in Figure \ref{survey_area_fig} because their clustering~(and cosmic variance) could not be accurately probed due to excessive shot noise. Here, we simply quantify the likelihood of their detection by determining the fraction of survey realizations within \texttt{BlueTides} that contain these bright outliers. Figure \ref{outlier_galaxies} shows the overall probability as a function of redshift for absolute UV magnitude thresholds ranging from $\sim-21$ to $\sim-25$. Note that $M_{UV}\sim-22$ corresponds to the magnitude of GNz11~\citep[hereafter O16]{2016ApJ...819..129O}. For GNz11 type galaxies~(red lines), the likelihood of detection is about $\sim4-5\%$ at $z\sim11\pm0.5$ for JADES medium survey. Due to the somewhat smaller volume for JADES-medium survey~($\sim3\times10^5~\mathrm{Mpc}^3$ at $z=11\pm0.5$) compared to volume of O16~($\sim1.2\times10^6~\mathrm{Mpc}^3$), the detection probability of GNz11 like galaxies in the JWST medium / deep surveys is lower than that within the O16 volume, i.e.  $\sim13\%$ according to \texttt{BlueTides}~\citep{2016MNRAS.461L..51W}. If we look at objects $1-2$ magnitudes brighter than GNz11, $M_{UV}\sim-23$~($M_{UV}\sim-24$) galaxy has up to $\sim10\%$ chance of getting detected for redshifts up to $\sim9$~($\sim8$) for the JADES medium survey. For the JADES deep survey, the corresponding probabilities fall by about a factor of $\sim5$.

\subsubsection{Lensed volumes}
`Lensed' surveys are obtained by looking at gravitationally lensed backgrounds of massive clusters~(e.g. Abell 2744, MACSJ0416.1-2403). Examples from current surveys include the Hubble Frontier Fields~\citep{2017AAS...23031613K}. The magnification due to lensing makes it possible to detect objects 2-4 magnitudes deeper than the limiting magnitude~(in the absence of lensing). 

\begin{figure*}
\includegraphics[width=16cm]{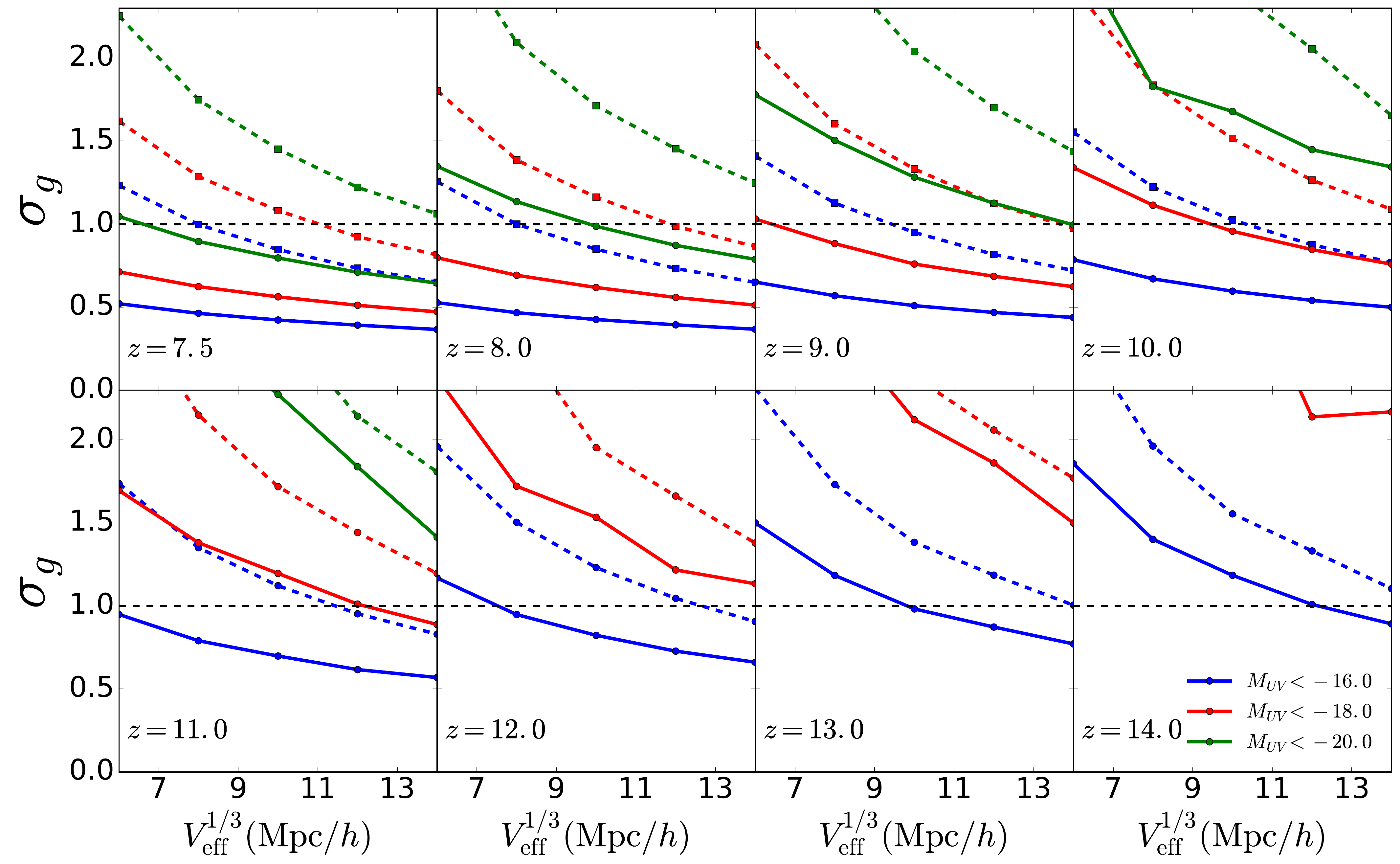}
\caption{\textbf{Cosmic variance in lensed surveys:} $\sigma_{\mathrm{galaxy}}$ is the cosmic variance as a function of volumes for lensed surveys for various $M_{UV}$ thresholds. We consider a range of volumes~($V_{\mathrm{eff}}$) based on effective volumes of HST Frontier fields surveys~\citep{2017AAS...23031613K} as computed in Figure 10 of \protect\cite{Livermore2017}. The solid and dashed lines have the same total survey volume, but have different geometries. The solid lines correspond to pencil-beam like geometries with $\Delta z=1$. The dashed lines, on the other hand, correspond to cubic geometries.} 
\label{sigma_lensed_field}
\end{figure*}

In order to estimate the cosmic variance for these lensed surveys, we consider simulation sub-volumes over the range of $~(\sim6-15~\mathrm{Mpc}/h)^3$, based on the  effective volume~($V_{\mathrm{eff}}$) estimates made by \cite{Livermore2017} using lensing models~\citep[and references therein]{Brada__2009,2015MNRAS.452.1437J,2016ApJ...819..114K}. Figure \ref{sigma_lensed_field} shows the cosmic variance as a function of volume for redshifts 7.5-14 up to UV magnitudes of $-16$. We do not focus on galaxies fainter than $M_{UV}\sim-16$ as their statistics may be affected by limited particle resolution; we however note that current and future lensed surveys can reach upto $\sim2-3$ magnitudes fainter. We present the results for two different geometries~(at fixed volume) i.e. ``pencil beam" like geometries assuming a redshift uncertainty of $\Delta z=1$~(solid lines), and cubic geometries~(dashed lines). We see that the cosmic variance is $\sim40\%$ or higher across the entire range of magnitudes and redshifts. Additionally, there are also conditions~(high enough luminosity, redshift or small enough volume) when the cosmic variance can exceed $100\%$, in which case the measurements are of limited value for providing constraints on the underlying physics. We therefore identify regimes under which the cosmic variance is contained within $\sim100\%$.  We primarily focus on $M_{UV}<-16$~(blue line) and $M_{UV}<-18$~(red line) since these surveys are primarily targeting the faint end of the luminosity function. We shall first summarize the results for the pencil beam geometries: for $M_{UV}<-16$, the cosmic variance is below $\sim100\%$ for the entire range of effective volumes up to $z\sim11$. At higher redshifts, to keep the cosmic variance of $M_{UV}<-16$ galaxies below $100\%$, the volumes required are  $\gtrsim(8~\mathrm{Mpc}/h)^3$ for $z\sim12$, $\gtrsim(10~\mathrm{Mpc}/h)^3$ for $z\sim13$ and $\gtrsim12~(\mathrm{Mpc}/h)^3$ for $z\sim14$. Likewise, for $M_{UV}<-18$ galaxies, the cosmic variance is below $\sim100\%$ for the entire range of effective volumes up to $z\sim9$. At higher redshifts, the cosmic variance is kept below $100\%$ at volumes $\gtrsim (9~\mathrm{Mpc}/h)^3$ for $z\sim10$ and $\gtrsim (12~\mathrm{Mpc}/h)^3$  for $z\sim11$. At $z>12$, the cosmic variance is $>100\%$ for the entire range of volumes presented. We now compare the results for the pencil beam vs. cubic geometries~(solid vs dashed lines in Figure \ref{sigma_lensed_field}); we find the cubical volumes have higher cosmic variance up to factors of $2-3$ at fixed $V_{\mathrm{eff}}$. As we increase the effective volume, the difference between the estimates for pencil beam vs. cubic geometries decreases. This is expected because at smaller survey volumes, the cubical geometries are expected to enclose extreme overdensities/ underdensities of galaxies, which is not expected in pencil beam geometries due to their line of sight dimensions ($\gtrsim100~\mathrm{Mpc}/h$) being significantly larger than the typical galaxy clustering scales~($\sim5-10~\mathrm{Mpc}/h$).



\begin{figure*}
\includegraphics[width=13cm]{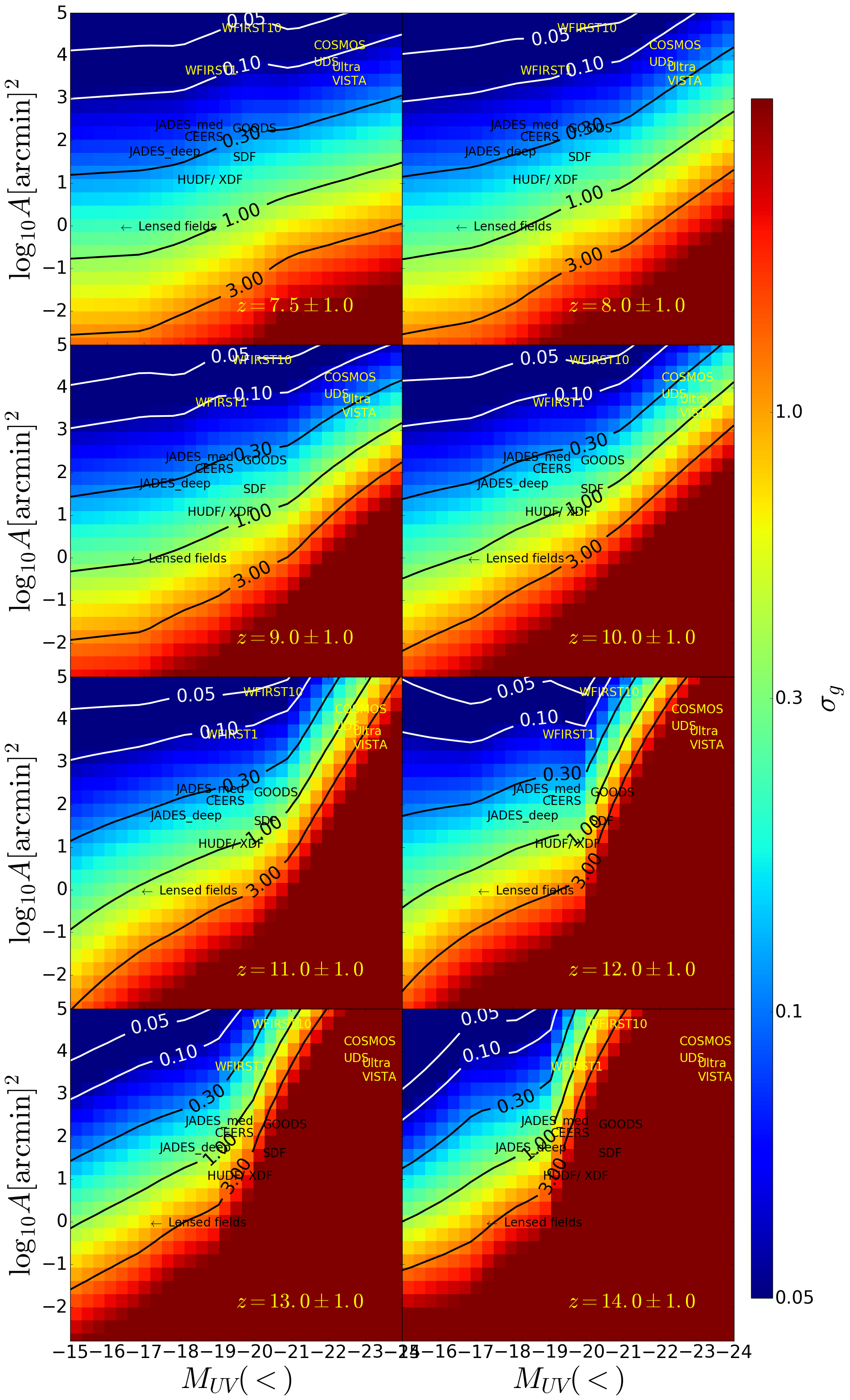}
\caption{The color map shows the cosmic variance as a function of threshold $M_{UV}$ magnitude and survey area $A$ as calculated by \href{https://github.com/akbhowmi/CV_AT_COSMIC_DAWN}{\texttt{CV_AT_COSMIC_DAWN}}. The solid black lines show contours representing $\sigma_g\sim0.1,0.3,1,3,10$. We show upcoming~(JWST, WFIRST) and current~(HUDF, SDF, CANDELS) surveys at various points on the plane positioned approximately by their survey area and limiting $H$-band magnitude~(converted to $M_{UV}$). We also show upcoming~(JWST lensed) and current~(Hubble and Subaru Frontier fields) lensed surveys collectively as `Lensed surveys'. The left arrow indicates that the limiting magnitudes for the lensed surveys may be 3-4 magnitudes fainter than the faintest galaxies \texttt{BlueTides} can probe.}
\label{sigma_surveys}
\end{figure*}


\subsection{Constructing \href{https://github.com/akbhowmi/CV_AT_COSMIC_DAWN}{\textrm{\texttt{CV_AT_COSMIC_DAWN}}}: A cosmic variance estimator for $z>7$ galaxies}
We use the results of the previous two sections to construct a cosmic variance calculator \href{https://github.com/akbhowmi/CV_AT_COSMIC_DAWN}{\texttt{CV_AT_COSMIC_DAWN}}~(all occurrences of `\texttt{CV_AT_COSMIC_DAWN}' are hyperlinks to the github repository) for $z>7$. In particular, \href{https://github.com/akbhowmi/CV_AT_COSMIC_DAWN}{\texttt{CV_AT_COSMIC_DAWN}} uses the fitting results summarized in Table \ref{power_law_fits_table} and Eq. \ref{redshift_width_variation_eqn} to compute cosmic variances for $M_{UV}$ thresholds and redshifts listed in Table \ref{power_law_fits_table}. For the redshifts and $M_{UV}$ thresholds which lie in between those listed in Table \ref{power_law_fits_table}, we use linear interpolation to estimate the cosmic variance. Cosmic variance estimates made using \href{https://github.com/akbhowmi/CV_AT_COSMIC_DAWN}{\texttt{CV_AT_COSMIC_DAWN}} are shown as dotted lines in Figure \ref{stellar_mass_threshold}.     

\begin{table*}
\centering
\begin{tabular}{|c|c|c|c|c|}
Survey & Instrument&Area  & $H~(<)$ & Reference\\
\hline
WFIRST10 & WFIRST & $10~\mathrm{deg}^2$ & $\sim27.5$ &\href{https://wfirst.gsfc.nasa.gov/science/WFIRSTScienceSheetFINAL.pdf}{WFIRST Science Sheet} \\
WFIRST1 & WFIRST & $1~\mathrm{deg}^2$ & $\sim28.5$ &\href{https://wfirst.gsfc.nasa.gov/science/WFIRSTScienceSheetFINAL.pdf}{WFIRST Science Sheet}\\
ultraVISTA & VISTA & $1~\mathrm{deg}^2$ & $\sim28.5$ &\cite{refId0}\\
JADES-medium & JWST & $\sim 190~\mathrm{arcmin}^2$ & $\sim29.7$& \href{https://issues.cosmos.esa.int/jwst-nirspecwiki/display/PUBLIC/Overview}{JADES survey overview}\\
JADES-deep & JWST & $\sim46~\mathrm{arcmin}^2$ & $\sim30.6$&\href{https://issues.cosmos.esa.int/jwst-nirspecwiki/display/PUBLIC/Overview}{JADES survey overviews}\\
CEERS & JWST & $\sim100~\mathrm{arcmin}^2$ & $\sim29$& \cite{2017jwst.prop.1345F}\\
HUDF & HST & $\sim10~\mathrm{arcmin}^2$ & $\sim27$&\cite{2015AJ....150...31R}\\
GOODS & HST & $\sim160~\mathrm{arcmin}^2$ & $\sim27.7$&\cite{2011ApJS..197...35G}\\
COSMOS & HST & $\sim2~\mathrm{deg}^2$ & $\sim25.5$&\cite{2011ApJS..197...35G}\\
UDS & HST & $\sim0.8~\mathrm{deg}^2$ & $\sim25$&\cite{2011ApJS..197...35G}\\
SDF & HSC & $\sim34.~\mathrm{arcmin}^2$ & $\sim27.5$&\cite{2004PASJ...56.1011K}\\
\hline
\end{tabular}
\caption{List of upcoming and current high redshift surveys using WFIRST, JWST, Hubble space telescope~(HST), Hyper Suprime Cam~(HSC) and Cosmic Assembly Near infrared Extra-gaLactic Survey~(CANDELS). HUDF refers to Hubble Ultra Deep Field and SDF refers to Subaru Deep Field. $H(<)$ is the detection limit in the $H$ band of WFIRST}
\label{list_of_surveys_table}
\end{table*}


We use \href{https://github.com/akbhowmi/CV_AT_COSMIC_DAWN}{\texttt{CV_AT_COSMIC_DAWN}} to summarize our results as a 2D color plot~(Figure \ref{sigma_surveys}) on the $A-M_{UV}(<)$ plane. We also present our estimates in terms of the apparent magnitude in Appendix \ref{appendix1}. The cosmic variance ranges from $\sigma_g\sim0.01$ to $\sigma_g\sim10$ and is represented by pixels colored as blue to red respectively. The solid black lines show contours representing $\sigma_g\sim0.1,0.3,1,3,10$. We show all the recent and upcoming surveys listed in Table \ref{list_of_surveys_table} as various points on the plane positioned approximately by their survey area and depth. 



\subsection{Implications for galaxy luminosity functions: Contribution of cosmic variance to total uncertainty}
\begin{figure*}
\includegraphics[width=16cm]{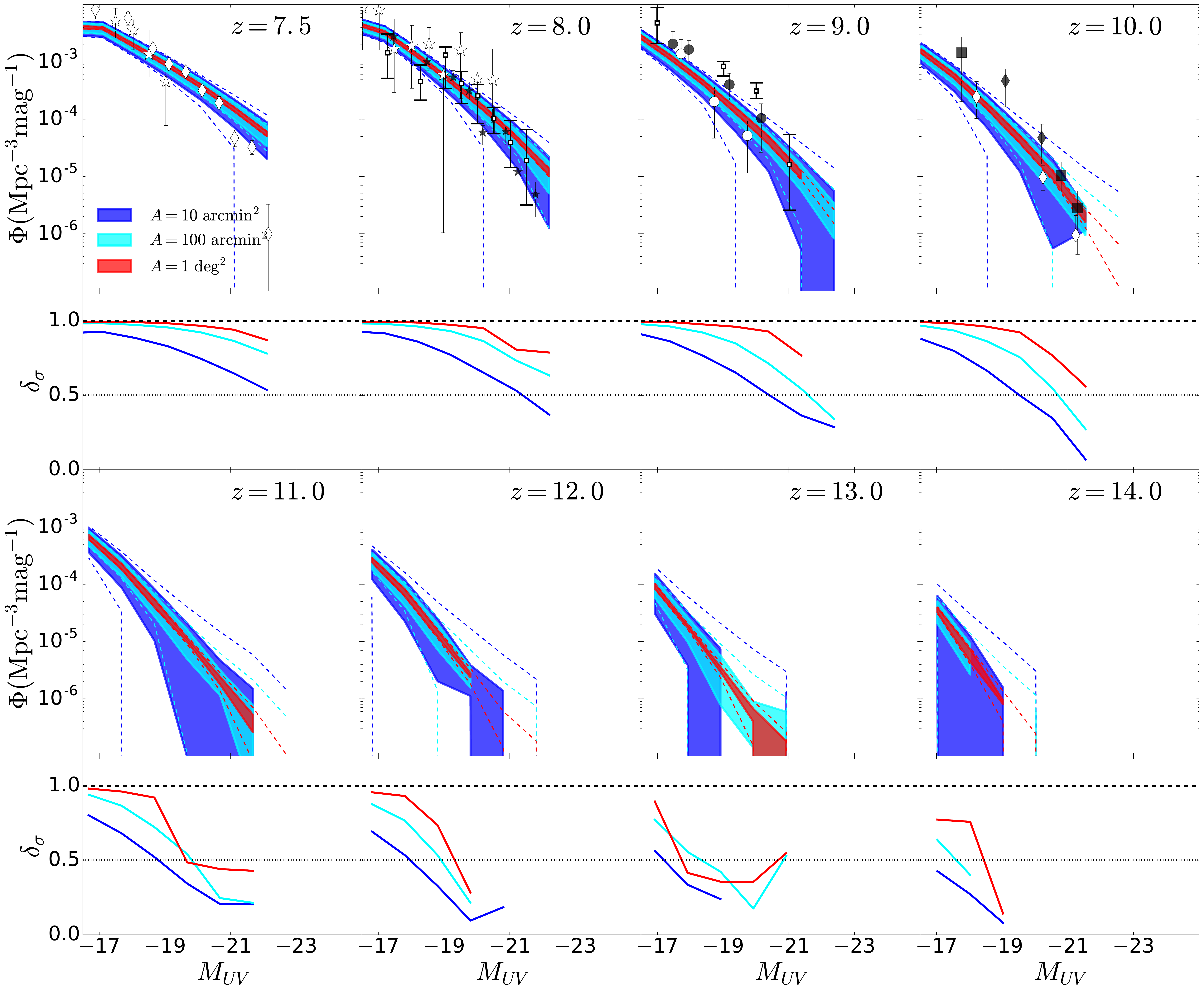}
\caption{\textbf{Top Panels:} $\Phi$ is the rest frame UV luminosity function. Different colors represent galaxies within simulation sub-volumes corresponding to different survey areas with $\Delta z=1$. For each color, the shaded region corresponds to uncertainty due to cosmic variance. For each color, the dashed lines are upper and lower limits representing the total field to field variance~(cosmic variance + Poisson variance). \textbf{Bottom Panels:} $\delta_\sigma$ is the ratio between the cosmic variance and the total field to field variance. Open stars~\citep{Livermore2017}, open squares~\citep{2018ApJ...854...73I}, open diamonds~\citep{2015ApJ...803...34B}, open circles~\citep{2012A&A...542L..31L}, filled stars~\citep{2015ApJ...803...34B}, filled diamonds~\citep{2016MNRAS.459.3812M}, filled squares~\citep{2018ApJS..237...12O} are observational measurements from current deep and lensed fields.}
\label{luminosity_function}
\end{figure*}
We now study the impact of cosmic variance on the galaxy luminosity function. In figure \ref{luminosity_function}, we compute the luminosity function and the associated cosmic variance and total=cosmic+Poisson variance for various survey areas. The open and black points show current observational constraints. The cosmic variance, shown by the shaded regions, reflects the trends seen in Figure \ref{stellar_mass_threshold}, and is broadly consistent with uncertainties in observational measurements which typically include cosmic variance estimates.


The bottom panels show the  fraction~($\delta_\sigma$) of the total uncertainty that is contributed by cosmic variance. For fixed magnitude, we see that as survey area decreases, $\delta_\sigma$ decreases. Likewise, for fixed survey area, we see that as galaxies become brighter, $\delta_\sigma$ decreases. This is expected since number counts decrease with decreasing survey area and with increasing luminosity, which increases the contribution from Poisson variance. Furthermore, we see that $\delta_\sigma>50\%$, implying that cosmic variance is the more dominant contribution to the overall uncertainty as compared to Poisson variance,~the only exceptions being samples with very small~($\lesssim10$) number counts~(as also seen in Figure \ref{cosmic_vs_poisson}).  
\section{Possible uncertainties in the cosmic variance estimates}
Our cosmic variance estimates are subject to uncertainties, particularly because the estimates are based on a single hydrodynamic simulation run with a fixed cosmology and galaxy formation modeling. The cosmic variance estimates depend on cosmology due to its effect on the halo bias and matter clustering, as well as the comoving survey volume. For instance, between WMAP~\citep{2013ApJS..208...19H} and PLANCK~\citep{2016A&A...594A..13P} cosmologies, the comoving survey volume changes by $\sim15~\%$; the matter clustering changes by $\sim4-10~\%$~(depending on the length scale) and the halo bias~(based on the \cite{2010ApJ...724..878T} model) changes by $\sim0.5-3~\%$~(depending on the halo mass scale). Adding these contributions up, we can overall expect a difference of $\sim25-30~\%$ between cosmic variances $\sigma_g$ predicted by the WMAP and PLANCK cosmologies. Additionally, uncertainties in the galaxy formation physics can also affect our cosmic variance estimates. In particular, a given sample of galaxies can populate haloes of different masses in different recipes of galaxy formation, thereby affecting the clustering amplitudes. For example, if the star formation within a galaxy sample is extremely `bursty' or `episodic', they may reside within a relatively small fraction of lower mass~(more abundant) haloes, compared to a model that does not lead to bursty star formation. This will lead to lower clustering amplitude~(for a fixed number density or luminosity function). Finally, we make several approximations in computing the cosmic variance: we use cuboidal volumes through the box with fixed transverse extent, rather than lightcones, and do not include the time evolution across the redshift interval. We expect that the errors due to these approximations will not significantly affect our predictions.



\section{Summary and Conclusions}
\label{summary_sec}
In this work, we used the recent \texttt{BlueTides} simulation to estimate the cosmic variance for $z>7$ galaxies to be detected by the planned deep surveys of JWST and WFIRST. Cosmic variance is expected to be a significant, potentially dominant source of uncertainty given the exceptionally strong clustering power~(galaxy bias $\gtrsim6$) of these galaxies seen in recent observations. We express the cosmic variance as an integral of the two-point correlation function over the survey volume, as commonly  done in the literature \citep{1980lssu.book.....P,2011ApJ...731..113M}.      


The resolution and volume enables \texttt{BlueTides} to probe the large scale bias, and therefore the cosmic variance of $z>7$ galaxies with $M_{UV}\sim-16$ to $M_{UV}\sim-22$ over survey areas $\sim0.1~\mathrm{arcmin}^2$ to $\sim 10~\mathrm{deg}^2$. Within this regime, the cosmic variance has a power law dependence on survey volume~(with exponent $\sim -0.25$ to $-0.45$). More luminous galaxies have larger cosmic variance than faint galaxies. 

The above trends can be put in the context of upcoming deep surveys. The largest planned deep survey will naturally suffer from the least amount of cosmic variance; this corresponds to the $10~\mathrm{deg}^2$ field of WFIRST, which will have a cosmic variance ranging from $\sim3-10\%$, except for $M_{UV}<-22$ galaxies at $z>12$ where the cosmic variance can exceed $\sim10\%$. Upcoming JWST medium/ deep surveys~(up to areas of $100~\mathrm{arcmin}^2$) will have a cosmic variance of about $20-50\%$ for $M_{UV}$ between $-16$ to $-20$. At the other end, the smallest surveys are the lensed surveys~(Hubble Frontier fields) and are most susceptible to cosmic variance. They have cosmic variance $\gtrsim40\%$ over the entire range of magnitudes and redshifts. These are the only existing surveys that can probe the faint~($M_{UV}$ thresholds between $\sim-13$ to $-16$) end of the luminosity function. In order for these measurements to  provide useful constraints~(e.g. on the nature of dark matter), the cosmic variance must be contained within $100\%$. For $M_{UV}$ thresholds up to $-16$, the cosmic variance is within $100\%$ for $z\sim7-11$ for the entire range of effective volumes between $\sim6-14~(\mathrm{Mpc}/h)^3$. At higher redshifts, effective volumes of $\gtrsim 8~(\mathrm{Mpc}/h)^3$ and $\gtrsim 12~(\mathrm{Mpc}/h)^3$ at $z\sim12$ and $z\sim14$ respectively, to keep the cosmic variance within $100\%$. 

Lastly, we study the impact of cosmic variance on the luminosity function and estimate the contribution of cosmic variance to the total uncertainty. We find that across all redshifts and magnitude bins~(with the exception of the most luminous bins with number counts $\lesssim10$ objects), cosmic variance is the more dominant component of the uncertainty, as compared to Poisson variance.  

We capture our results in the form of simple fitting functions and encode them in an online cosmic variance calculator (\href{https://github.com/akbhowmi/CV_AT_COSMIC_DAWN}{\texttt{CV_AT_COSMIC_DAWN}}) which we publicly release.

\appendix
\section{Presenting cosmic variance in terms of telescope survey parameters}
\label{appendix1}
\begin{figure*}
\includegraphics[width=14cm]{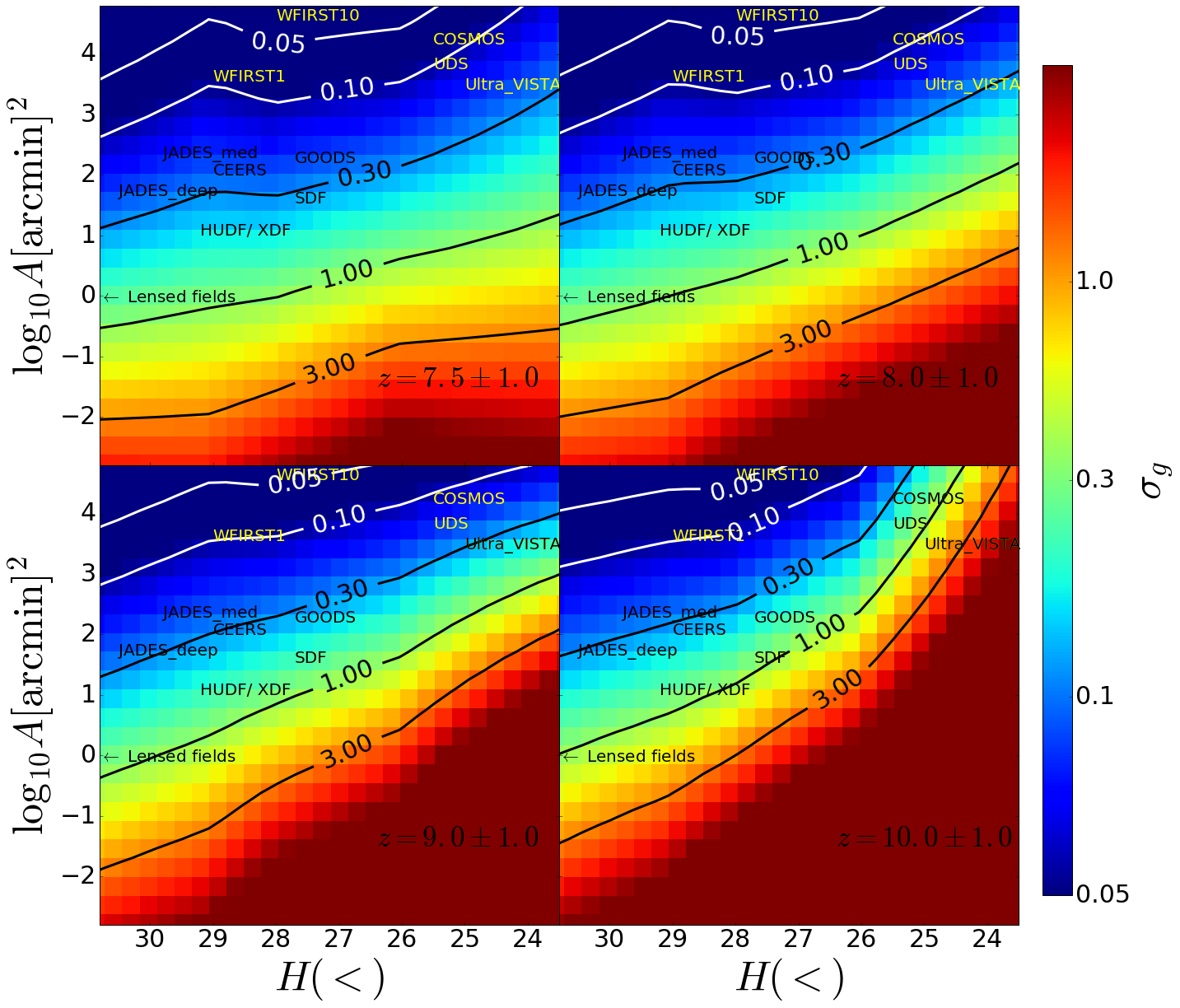}
\caption{The color map shows the cosmic variance as a function of $H$ band~(within WFIRST) magnitude and survey area $A$ as calculated by \href{https://github.com/akbhowmi/CV_AT_COSMIC_DAWN}{\texttt{CV_AT_COSMIC_DAWN}}. The solid black lines show contours representing $\sigma_g\sim0.1,0.3,1,3,10$. We show upcoming~(JWST, WFIRST) and current~(HUDF, SDF, CANDELS) surveys at various points on the plane positioned approximately by their survey area and limiting $H$-band magnitude. We also show upcoming~(JWST lensed) and current~(Hubble and Subaru Frontier fields) lensed surveys collectively as `Lensed surveys'. The left arrow indicates that the limiting magnitudes for the lensed surveys may be 3-4 magnitudes fainter than the faintest galaxies \texttt{BlueTides} can probe.}
\label{sigma_surveys_Hband}
\end{figure*}

In addition to presenting cosmic variance as a function of the rest frame intrinsic UV magnitude, it is also useful~(for observers in particular) to present our estimates directly in terms of the apparent magnitude, which is fixed for a given survey. We therefore, in Figure \ref{sigma_surveys_Hband}, also present our estimates in terms of the apparent magnitude. We choose the $H$ band magnitude of WFIRST, and present our estimates up to redshift 10.    

\section*{Acknowledgements}
We acknowledge funding from NSF ACI-1614853, NSF AST1517593, NSF AST-1616168, NSF AST-1716131, NASA ATP
NNX17AK56G and NASA ATP 17-0123, and the BLUEWATERS
PAID program. NSF:- National Science Foundation, ACI:- Division
of Advanced Cyber Infrastructure, AST:- Division of Astronomical
Sciences, NASA:- National Aeronautics and Space Administration,
ATP:- Astrophysics Theory Program. The BlueTides simulation
was run on the BLUEWATERS facility at the National Center for
Supercomputing Applications.
\bibliography{example}
\end{document}